\UseRawInputEncoding
\documentclass[%
 reprint,
superscriptaddress,
amsmath,amssymb,
aps,
prb,
floatfix,
longbibliography
]{revtex4-2}

\usepackage{graphicx}
\usepackage{dcolumn}
\usepackage{bm}
\usepackage{hyperref}
\usepackage{natbib}

\usepackage{xcolor}
\definecolor{colorblue}{RGB}{4,4,236}
\hypersetup{colorlinks=true, linkcolor=colorblue, citecolor=colorblue, urlcolor=colorblue}

\begin{document}

\preprint{APS/123-QED}

\title{Dispersion Characteristics of a Glide-Symmetric Square Patch Metamaterial with Giant Anisotropy}

\author{Jim A. Enriquez}
\affiliation{School of Physics and Engineering, ITMO University, 197101, Saint Petersburg, Russia}
\affiliation{National University of Colombia, Department of Physics, 111321, Bogota, Colombia}
\author{Eugene Koreshin}
\affiliation{School of Physics and Engineering, ITMO University, 197101, Saint Petersburg, Russia}
\author{Juan P. Del Risco}
\affiliation{National University of Colombia, Department of Physics, 111321, Bogota, Colombia}
\author{Pavel A. Belov}
\affiliation{School of Physics and Engineering, ITMO University, 197101, Saint Petersburg, Russia}
\affiliation{Qingdao Innovation and Development Center, Harbin Engineering University, Qingdao 266000, Shandong, China}
\affiliation{School of Engineering, New Uzbekistan University, Movarounnahr str. 1, 100000, Tashkent, Uzbekistan}
\author{Juan D. Baena}
\affiliation{National University of Colombia, Department of Physics, 111321, Bogota, Colombia}
\date{\today}

\begin{abstract}
This paper investigates the dispersion characteristics of a highly anisotropic metamaterial comprised of metal square patches arranged in a glide symmetry pattern and submerged in vacuum. Theoretical formulas are proposed to describe the electromagnetic tensors of a corresponding uniaxial effective medium with dielectric and magnetic responses. In addition, this work employs theoretical analysis and numerical simulations to examine the interaction between the metamaterial and electromagnetic waves across a broad spectral range. Band diagrams and isofrequency contours show good agreement between theoretical and numerical results for low frequencies and certain directions of propagation at higher frequencies. The ease of designing the metamaterial structure for various applications is facilitated by the derived theoretical formulas, which enable accurate prediction of the electromagnetic response across a wide range of frequencies based on geometric parameters. 

\end{abstract}

\maketitle


\section{Introduction}
Metamaterials are artificial materials with electromagnetic properties beyond those found in natural materials. Metamaterials often exhibit anisotropy, meaning that material responses depend on the orientation of the electromagnetic field. Studies have revealed exotic electromagnetic properties in anisotropic metamaterials such as negative-refraction \cite{poddubny_hyperbolic_2013}, near-zero parameters \cite{zhang_effective_2015}, and manipulation of polarization states in reflection \cite{hao_manipulating_2007}.

Symmetries, together with an adequate selection of parameters and constituents of metamaterial structures, enable the design of materials with particular electromagnetic wave propagation properties. A symmetry that has been exploited to attain exceptional electromagnetic responses is a higher symmetry known as glide symmetry. Structures exhibiting glide symmetry possess invariance under the combined operations of a reflection across a glide plane and a subsequent translation by half a period along each direction parallel to the glide plane \cite{1450933}. It has been shown that with an adequate selection of parameters, metamaterials based on glide symmetry allow to reduce the dispersion \cite{quevedo-teruel_periodic_2020, quevedo-teruel_benefits_2021} and increase the effective refractive index \cite{ebrahimpouri_ultrawideband_2019, arnberg_high_2020}.

\begin{figure}[ht]
\includegraphics[width=1\columnwidth]{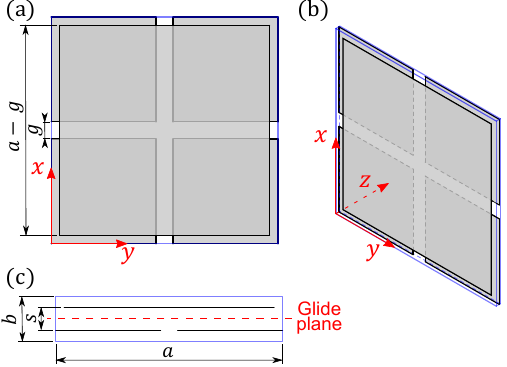}%
\caption{Supercell of the metallic square patch metamaterial, where the spatial period along the $z$-axis is $b$, the spatial periods along the $x$- and $y$-axes are both $a$, and the distance between adjacent planes of patches is $s=b/2$. (a) Frontal view, (b) isometric view, and (c) bottom view. The scale of the ratio $b/a$ is augmented by a factor of 8 in Fig. (c) to improve visualisation.}\label{fig_1}
\end{figure}

A unique anisotropic metamaterial based on glide symmetry consist of an array of square metal patches that are repeated with a period $a$ along the directions parallel to the patches (transverse directions), and a period $b$ along the direction perpendicular to the patches (axial direction), as shown Fig. \ref{fig_1} for a supercell of the structure. The distance between adjacent planes of patches is half of the axial period, $s=b/2$. The glide plane is located at the center of the unit cell along the z-axis. To replicate the adjacent plane, each patch undergoes a reflection about the glide plane followed by a translation of half the transverse period, $a/2$, along the $x$ and $y$ directions. With a much larger period along the transversal directions compared to the axial direction  ($a>>b$), a high dielectric permittivity uniaxial material is obtained \cite{chang_broadband_2016, barzegar-parizi_calculation_2015}. Taking advantage of the unique characteristics of the metamaterial, it has been proposed as a component of practical applications related with microwave engineering \cite{ma_artificial_2008, takahagi_high-gain_2011} and magnetic resonance imaging \cite{vorobyev_artificial_2020}.  

Effective medium models have been proposed to study the dispersion characteristics of the square patch metamaterial along the axial direction, considering a uniform electric charge across the metal patches \cite{chang_broadband_2016}. Furthermore, analytical models have been put forth to realize the scattering parameters of plane waves impinging structures based on layers with non aligned metal patches \cite{ cavallo_analytical_2017, 8457250}. However, the magnetic behavior of the structure and the frequency range of validity for existing effective medium models are not well established yet. In this article, we present theoretical formulas that predict the dispersion properties of the square patch metamaterial depending on the geometrical parameters of the structure. The theoretical results are compared with numerical ones to check the scope of validity of the proposed effective medium. Dispersion diagrams and isofrequency contours of propagating modes supported by the structure are analyzed. In addition, the electromagnetic parameters are theoretically and numerically retrieved for low frequencies. Our results show that the effective medium approach is valid for low frequencies and for high frequencies along certain directions of propagation.

\section{Theory}
The theoretical model is based on a quasi-static approach including both electrostatic and magnetostatic studies \cite{Baena_boradband_uniaxial2020}. Given the subwavelength dimensions of the square patch medium, it can be effectively described as an effective medium for low frequencies \cite{tretyakov2003analytical}. For simplicity, the host medium is considered as vacuum, and the metallic patches are considered made of Perfect Electric Conductor (PEC). Bianisotropic effects are discarded because of the presence of a center of symmetry in the unit cell and non-linear effects are not considered. Consequently, linear constitutive relations are assumed. Moreover, due to mirror symmetries of the unit cell, both the permittivity and permeability tensors are diagonal. Therefore, the constitutive relations can be written as, 
\begin{equation}\label{eq_1}
                    \mathbf{D}=\epsilon_0\begin{pmatrix}
                    \epsilon_t & 0 & 0\\
                    0 & \epsilon_t & 0\\
                    0 & 0 & 1
                    \end{pmatrix}\mathbf{E},
                    \end{equation}
\begin{equation}\label{eq_2}
                    \mathbf{B}=\mu_0\begin{pmatrix}
                    1 & 0 & 0\\
                    0 & 1 & 0\\
                    0 & 0 & \mu_z
                    \end{pmatrix}\mathbf{H},
\end{equation}
in which the unknown parameters are the relative transverse permittivity ($\epsilon_t$) and the relative axial permeability ($\mu_z$). For identifying the components of the electromagnetic tensors it is considered that the field components $E_z$, $B_x$, and $B_y$ do not interact with the conducting patches, which are assumed infinitesimally thin. Then $\epsilon_z =\mu_x=\mu_y=1$. In addition, the structure is symmetric under a $90^{\circ}$ rotation around the $z$-axis, implying $\epsilon_x  = \epsilon_y = \epsilon_t$.

\subsection{Transverse relative permittivity}

\begin{figure}[ht]
\includegraphics[width=1\columnwidth]{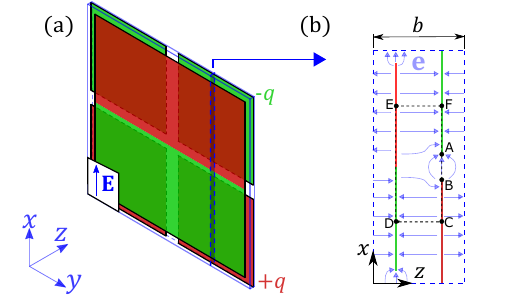}%
\caption{Electric response of the square patch metamaterial to a macroscopic electric field oriented along the $x$-axis. (a) Induced charge and  (b) microscopic electric field ($\mathbf{e}$) distribution within the unit cell. The scale of the ratio $b/a$ is augmented by a factor of 16 in Fig. (b) to improve visualisation.}\label{fig_2}
\end{figure}

It is assumed that both the macroscopic electric field $\mathbf{E}$ and the polarization $\mathbf{P}$ are oriented along the $x$-axis, considering that the macroscopic electric field is understood as the average electric field along the unit cell. The electric responses within the unit cell due to this $x$-oriented $\mathbf{E}$ are illustrated in Fig. \ref{fig_2}. The electric field along $x$ within the structure is only feasible along the air gaps, leading to a charge separation within the patches: one half becomes positive, and the other half becomes negative. This glide-symmetric structure enables a higher effective permittivity compared to non-glide-symmetric patches, in which the charge across the patches is uniform \cite{cavallo_analytical_2017}.

From the usual constitutive relation connecting the electric vectors, 

\begin{equation}
    D_x=\epsilon_0\ E_x+P_x=\epsilon_0\ \epsilon_t E_x,
\end{equation}
the transverse permittivity can be obtained,

\begin{equation}\label{eq: et1}
    \epsilon_t=1+ \frac{P_x}{\epsilon_0\ E_x\ }.
\end{equation}
Therefore, it is needed to analyze $P_x$ and $E_x$ to estimate the transverse permittivity.

To analyze $E_x$, the microscopic electric field $\mathbf{e}$ is used, which is the electric field in each point of the unit cell (see Fig. \hyperref[fig_2]{2b}). Following a vertical line passing through the points F, A, B, and C marked in Fig. \hyperref[fig_2]{2b}, it is noted that due to the boundary conditions in PEC, it is possible to obtain a non-zero electric field only in the gap of vacuum, so that,

\begin{equation}\label{eq:Exex}
    E_x a=-\int_A^Be_x\:dx.
\end{equation}
Conversely, applying Faraday's Law in the quasi-static regime along the closed integration path ABCDEFA depicted in Fig. \hyperref[fig_2]{2b},

\begin{equation}\label{eq:epath}
    \oint \mathbf{e}\cdot \:d\mathbf{r}=\int_A^B e_x \:dx + \int_C^D e_z\:dz + \int_E^F e_z \:dz \approx 0.
\end{equation}
Utilizing Eqs. \eqref{eq:Exex} and \eqref{eq:epath}, and assuming a uniform and equivalent $|e_z|$ along paths CD and EF,

\begin{equation}\label{eq:Ex}
    E_x=\frac{2}{a}\int_C^De_z\: dz \approx \frac{b}{2a}\frac{\sigma}{\epsilon_0},
\end{equation}
where $\sigma$ is the uniform surface charge density in the half of a patch.

The polarization ($P_x$) is obtained by calculating the dipole moment due to two quarters of patches divided by the quarter of volume of the  supercell since there are two patches in the cell,

\begin{equation}\label{eq:Px}
    P_x=\frac{p_x}{ba^2/4}=\frac{qa/2}{ba^2/4}=\frac{2\sigma(\tfrac{a}{2}-g)^2}{ba},
\end{equation}
where $(\tfrac{a}{2}-g)^2$ represents the overlapping area between quarters of patches in adjacent layers. Substituting Eqs. \eqref{eq:Ex}, and \eqref{eq:Px} into Eq. \eqref{eq: et1} yields, 

\begin{equation}\label{eq_3ep1}
    \epsilon_t=1+\left(\frac{a-2g}{b}\right)^2.
\end{equation}
Including edge effects of the electric charge, the polarization is corrected as,
\begin{equation}\label{eq_px_c}
    P_x'=P_x+\Delta P=\frac{2\sigma(\tfrac{a}{2}-g)^2}{ba}+\frac{16\lambda(\tfrac{a}{2}-g)}{ba}.
\end{equation}
where the last term comes from the charge density along the edges of quarter of patches. $\lambda$ is a linear charge density, which is derived in Supplemental Material \cite{SM} \nocite{Pozar:882338}, $\lambda=0.4413 \sigma b/2$. Therefore, substituting Eqs. \eqref{eq:Ex} and \eqref{eq_px_c} into Eq. \eqref{eq: et1},

\begin{equation}\label{eq_3ep}
    \epsilon_t=1+\left(\frac{a-2g}{b}\right)^2+1.7652\frac{a-2g}{b}.
\end{equation}
It's noteworthy that as the axial period $b$ decreases, the transverse permittivity increases. This property can be leveraged to create exceptionally thin Fabry-Perot resonators \cite{7746420}.
\subsection{Axial relative permeability}

\begin{figure}[ht]
\includegraphics[width=1\columnwidth]{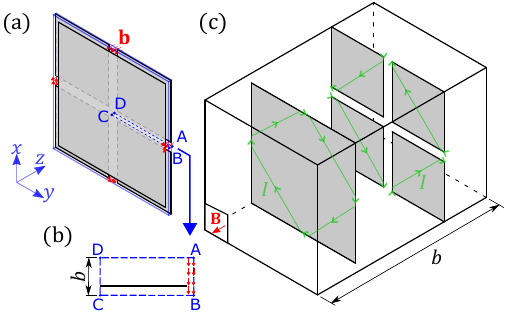}%
\caption{Magnetic response of the square patch metamaterial to a macroscopic magnetic flux density oriented along the $z$-axis. (a, b) Microscopic magnetic flux density ($\mathbf{b}$) distribution and (c) induced currents within the unit cell. The scale of the ratio $b/a$ is augmented by a factor of 60  in Fig. (b) and a factor of 8 in Fig. (c) to improve visualisation.}\label{fig_3}
\end{figure}

It is assumed that both the macroscopic magnetic flux density $\mathbf{B}$ and the magnetization $\mathbf{M}$ are oriented along the $z$-axis, considering that the macroscopic magnetic flux density is understood as the average magnetic flux density along the unit cell. The magnetic responses within the unit cell due to this $z$-oriented $\mathbf{B}$  are illustrated in Fig. \ref{fig_3}. Consequently, from the usual constitutive relation connecting the magnetic vectors,

\begin{equation}
    B_z=\mu_0 (H_z+M_z )=\mu_0 \mu_z H_z,
\end{equation}
the axial permeability can be obtained,

\begin{equation}\label{eq:muz}
    \mu_z=\frac{1}{1-\frac{\mu_0M_z}{B_z}}.
\end{equation}

Once $M_z$ and $B_z$ are identified, the axial permeability can be estimated. To analyze $B_z$, the microscopic magnetic flux density $\mathbf{b}$ is used, which is the magnetic flux density in each point of the unit cell. According to the boundary conditions for PEC, $b_z$ should be negligible in the regions bounded by parallel plates since they are tightly packed and $b_z = 0$ over the PEC. On the other hand, $b_z \neq 0$ along the canals formed by the crosses of orthogonal slots, see Fig. \hyperref[fig_3]{3a}. Therefore, applying Ampere's Law in the quasi-static regime along the closed integration path ABCDA depicted in Fig. \hyperref[fig_3]{3b} results in,
\begin{equation}\label{eq:bz}
    b_z=-\frac{\mu_0I}{b},
\end{equation}
where $I$ is the induced current in the metal patches, see Fig. \hyperref[fig_3]{3c}.

Equating the magnetic flux in the unit cell across the $xy$-plane from the macroscopic and microscopic points of view and using Eq. \eqref{eq:bz}, 
\begin{equation}\label{eq:Bz}
    B_z=\frac{b_z 2g^2}{a^2} =-\frac{g^2 \mu_0 I}{a^2 b/2},
\end{equation}

Regarding the magnetization, following the currents illustrated in Fig. \hyperref[fig_3]{3c} for one patch that assure a $b_z\neq0$ along the canals, 
\begin{equation}\label{eq:magnetization}
    M_z=\frac{m_z}{a^2 b/2}\approx\frac{AI}{a^2 b/2}=\frac{a^2/2-g^2}{a^2 b/2} I.
\end{equation}
Substituting Eqs. \eqref{eq:Bz} and \eqref{eq:magnetization} into Eq. \eqref{eq:muz},
\begin{equation}\label{eq_3mu1}
    \mu_z=\frac{2g^2}{a^2}.
\end{equation}
Including edge effects due to the microscopic magnetic flux density, the macroscopic magnetic flux density is corrected as,

\begin{equation}\label{eq_18}
    B_z'=B_z+\Delta B_z=\frac{1}{a^2}\left (2g^2 b_z +8g\lambda_m \right),
\end{equation}
where the last term comes from the magnetic flux along the edges of the canals within the unit cell. $\lambda_m$ is a linear density of magnetic flux, which is derived in Supplemental Material \cite{SM}, $\lambda_m=0.4413 b_zb/2$. Therefore, substituting Eqs. \eqref{eq:magnetization} and \eqref{eq_18} into Eq. \eqref{eq:muz},
\begin{equation}\label{eq_3mu}
    \mu_z=\frac{2g^2+1.7652gb}{a^2+1.7652gb}.
\end{equation}

In non-glide-symmetric patches, induced currents circulate along the patch perimeter, allowing magnetic flux to pass through the air gaps while blocking magnetic flux within the metal zones. As illustrated in Fig. \ref{fig_3}, the tightly packet glide-symmetric structure encourages current paths that avoid the patch edges. This configuration effectively blocks magnetic flux within the metal areas and allows magnetic flux to pass through the air canals. Consequently, the air canals in the glide-symmetric structure are narrower than those in non-glide-symmetric patches, leading to a lower effective permeability \cite{cavallo_analytical_2017}.

\subsection{Dispersion relations}
Using Maxwell's equations and considering plane wave propagation, the dispersion equations for the modes supported by the effective uniaxial medium are derived in accordance with \cite{depine_2006},

\begin{equation}\label{eq_4}
    \frac{\omega a}{2 \pi c}=\frac{a}{2\pi}\left(\frac{k_x^2+k_y^2}{\epsilon_t \mu_z}+\frac{k_z^2}{\epsilon_t}\right)^{1/2},
\end{equation}
\begin{equation}\label{eq_5}
    \frac{\omega a}{2 \pi c}=\frac{a}{2\pi}\left(k_x^2+k_y^2+\frac{k_z^2}{\epsilon_t}\right)^{1/2}.
\end{equation}
Eqs. \eqref{eq_4} and \eqref{eq_5} correspond to TE and TM modes, respectively. TE waves have an electric field perpendicular to the plane defined by the propagation vector and the optic axis ($\hat{z}$), while TM waves have a magnetic field perpendicular to this plane.

When waves propagate perpendicular to the metal patches, the material's behavior is primarily governed by the transverse permittivity ($\epsilon_t$). However, for grazing incidence, the transverse permeability ($\mu_z$) also becomes a significant factor. This type of metamaterial belongs to the class of uniaxial dielectric-magnetic media, established in the $1990_s$ \cite{lakhtakia1997time, weiglhofer1996new}. Recent studies have proposed the use of uniaxial dielectric-magnetic materials as a foundation for realizing angle-selective surfaces \cite{huang_angle-selective_2020} and impedance matching layers \cite{he_magnetoelectric_2018}.

The electric and magnetic effects within the metamaterial, as assumed in this section, give rise to capacitive and inductive effects, which can be modeled as capacitors and inductors. This modeling approach enables the derivation of theoretical formulas for the effective parameters. For instance, in \cite{7746420}, a similar expression to Eq. \eqref{eq_3ep1} was obtained by comparing an effective capacitor to the actual system of capacitors within the unit cell. Consequently, the methodology presented in this work could be valuable for understanding and identifying circuit models for other metastructures, such as mushroom-type metasurfaces \cite{798001}, which have been studied using glide symmetry to develop bandgap structures with increased operational bandwidth \cite{8972928}. 

\section{Numerical investigation}

\begin{figure}[ht]
{
  \includegraphics[width=1\columnwidth]{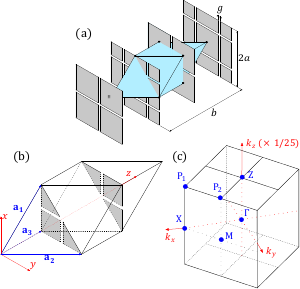}%
}
\caption{(a) Structure of the square patch metamaterial. (b) Primitive unit cell with translation vectors: $\mathbf{a_1}$, $\mathbf{a_2}$, and $\mathbf{a_3}$. (c) Brillouin zone and high-symmetry points with coordinates: $\Gamma=\left(0,0,0\right)^{\mathrm{T}}$, $\mathrm{X}=\left(\tfrac{2\pi}{a},0,0\right)^{\mathrm{T}}$, $\mathrm{M}=\left(\tfrac{\pi}{a},\tfrac{\pi}{a},0\right)^{\mathrm{T}}$, $\mathrm{Z}=\left(0,0,\tfrac{\pi}{b}+\tfrac{b\pi}{a^2}\right)^{\mathrm{T}}$, $\mathrm{P_1}=\left(\tfrac{2\pi}{a},0,\tfrac{\pi}{b}-\frac{b\pi}{a^2}\right)^{\mathrm{T}}$, $\mathrm{P_2}=\left(\tfrac{\pi}{a},\tfrac{\pi}{a},\tfrac{\pi}{b}\right)^{\mathrm{T}}$. The scale of the ratio $b/a$ is augmented by a factor of 120 in Figs. (a, b) and a factor of 25 in Fig. (c) to improve visualisation.}\label{fig_4}
\end{figure}

To illustrate the highly anisotropic response of the studied structure, we consider the following geometrical parameters: $b/a=0.025$, and $g/a=0.075$. Substituting  these values into Eqs. \eqref{eq_3ep} and \eqref{eq_3mu} yields in a relative transversal permittivity of $\epsilon_t=1217$ and a relative axial permeability of $\mu_z=0.01451$.

The periodic nature of the square patch metamaterial reassembles a body-centered tetragonal crystal system \cite{MEHL2017S1}. Therefore, a supercell of the structure consists in translation vectors $a \hat{x}$, $a \hat{y}$, and $b \hat{z}$ containing two patches, as illustrates Fig. \ref{fig_1}. In this study, we consider a primitive unit cell to numerically analyze the dispersion of the square patch metamaterial, as shown Fig. \hyperref[fig_4]{4b}. The election of a primitive unit cell to study a metamaterial is important to completely describe the dispersion characteristics of the structure and avoid redundant information, as was shown in \cite{sakhno_longitudinal_2021}, \cite{yang2023dispersion}. In addition, the selected unit cell fully exploits the glide symmetry of the structure, considering that each node in the cell is connected to adjacent points in different planes that are translated $\tfrac{a}{2}$ along both $x$ and $y$ directions (see Fig. \hyperref[fig_4]{4a}). Consequently, the components of the translation vectors of the primitive unit cell are given by, 
\begin{equation}\label{eq_6}
    \begin{split}
        \mathbf{a_1}=\left(\tfrac{a}{2},-\tfrac{a}{2},\tfrac{b}{2}\right)^{\mathrm{T}},\\
        \mathbf{a_2}=\left(-\tfrac{a}{2},\tfrac{a}{2},\tfrac{b}{2}\right)^{\mathrm{T}},\\
        \mathbf{a_3}=\left(-\tfrac{a}{2},-\tfrac{a}{2},\tfrac{b}{2}\right)^{\mathrm{T}}.
    \end{split}
\end{equation}

The Brillouin zone corresponding to the primitive unit cell, with the shape of a dodecahedron, is illustrated in Fig. \hyperref[fig_4]{4c}. Note that the Brillouin zone extends from $-2\pi/a$ to $2\pi/a$ along $k_x$. This is a consequence of the glide symmetry of the structure, which extends the periodicity in the reciprocal space along the directions parallel to the glide plane, according to the generalized Floquet theorem \cite{1450933}. The dispersion characteristics of the square patch metamaterial are numerically studied using the commercial software COMSOL MULTIPHYSICS, implementing periodic boundary conditions to the primitive unit cell, with $e^{i \mathbf{k}\cdot \mathbf{r}}$ spatial dependence. 

\subsection{Dispersion relations}
\begin{figure}[htb]
{%
  \includegraphics[width=1\columnwidth]{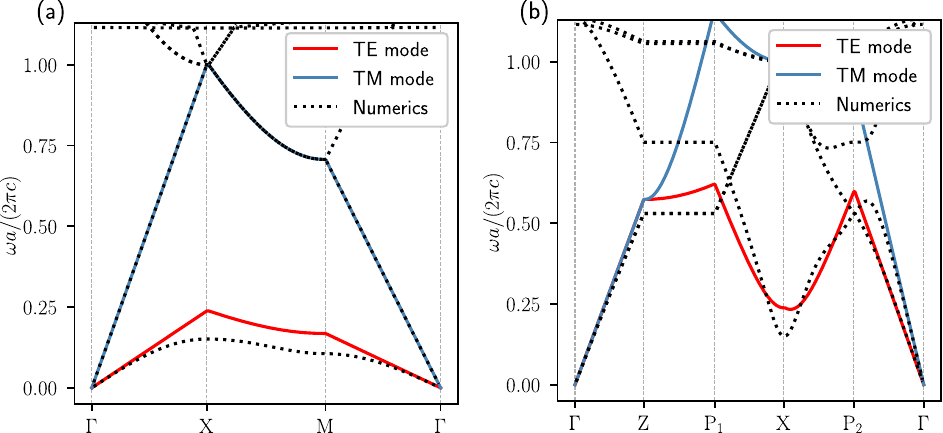}%
  }
\caption{Theoretical and numerical dispersion diagrams of the square patch metamaterial along paths involving the high symmetric points marked in Fig. \hyperref[fig_4]{4c}, (a) path $\mathrm{\Gamma XM\Gamma}$ and (b) path $\mathrm{\Gamma ZP_1XP_2\Gamma}$. TE mode and TM mode correspond to Eqs. \eqref{eq_4} and \eqref{eq_5}, respectively.}\label{fig_5}
\end{figure}

The dispersion relations are analyzed along paths involving the high symmetric points illustrated in Fig \hyperref[fig_4]{4c}. As a result, Fig. \ref{fig_5} illustrates theoretical and computational dispersion diagrams, in which TE mode and TM mode correspond to Eqs. \eqref{eq_4} and \eqref{eq_5}, respectively. 

The dispersion relations along path $\Gamma \mathrm{XM} \Gamma$ are illustrated in Fig. \hyperref[fig_5]{5a}. The TM mode is characterized by a lack of interaction with the metal patches, resulting in a perfect degree of similarity with numerical results and resembling the propagation of light in vacuum, as shown Eq. \eqref{eq_5} with $k_z=0$ . Conversely, the TE mode demonstrates good agreement for low frequencies, with a relative error of $7.1\%$ observed at the $\mathrm{X}/3$ point. The agreement between theoretical and computational results starts to degrade for regions close to the edges of the first Brillouin zone, which is a typical limitation of effective medium models \cite{Belov_homogenization_of_electromagnetic_crystals2005}. In this work, the relative error is calculated by comparing simulation results relative to theoretical predictions.

 \begin{figure}[ht]
\includegraphics[width=1\columnwidth]{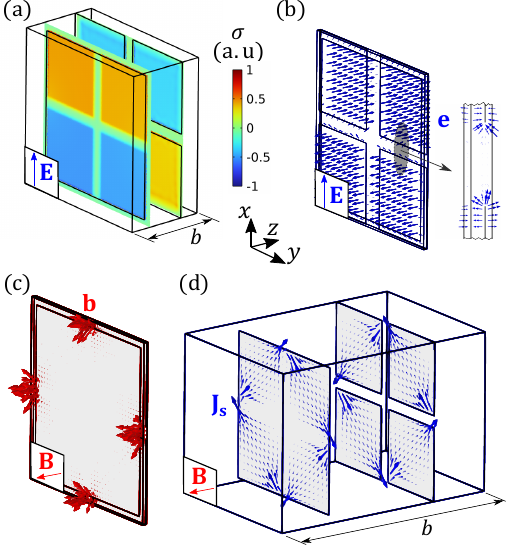}%
\caption{Computational results for the square patch metamaterial at frequency $\omega a/(2\pi c)=0.1$: (a, b) Electric response: surface charge density and microscopic electric field due to a plane wave with electric field along $x$. (c, d) Magnetic response: Microscopic magnetic flux density and surface current density due to a plane wave with magnetic field along $z$. The scale of the ratio $b/a$ is augmented by a factor of 20  in Fig. (a) and a factor of 60 in Fig. (d) to improve visualisation.}\label{fig_6}
\end{figure}

Including points with $k_z\neq0$, Fig. \hyperref[fig_5]{5b} illustrates the dispersion relations along path $\Gamma \mathrm{ZP_1XP_2} \Gamma$. There is a notable concurrence between theoretical and numerical results for both TE and TM modes below $\omega a/(2\pi c)=0.5$. In addition, along $k_z$ direction, which is the direction exhibiting a giant relative permittivity, the higher the frequency, the higher the relative error, with a maximum value of $8.3\%$ observed at $\mathrm{Z}$ point.

The agreement between theoretical and numerical results for the dispersion relations is further supported by the computational validation of the assumed electric and magnetic responses of the metamaterial. To this end, a plane wave with frequency $\omega a/(2\pi c)=0.1$, propagation vector along $y$-axis, electric field polarized along $x$-axis, and magnetic field polarized along $z$-axis  is impinged upon a supercell of the structure, as shown in Fig. \ref{fig_6}. The results corroborate the theoretical electric responses in Fig. \ref{fig_2} and the theoretical magnetic responses in Fig. \ref{fig_3}. As observed in Fig. \hyperref[fig_6]{6a}, the surface charge density vanishes at the center of the patch in both $x$ and $y$ directions. While this detail was not accounted for in the theoretical Fig. \ref{fig_2}, it does not impact the results since the effective permittivity depends on the overlapping area between patches in adjacent layers.
\begin{figure}[ht]
{%
\includegraphics[width=1\columnwidth]{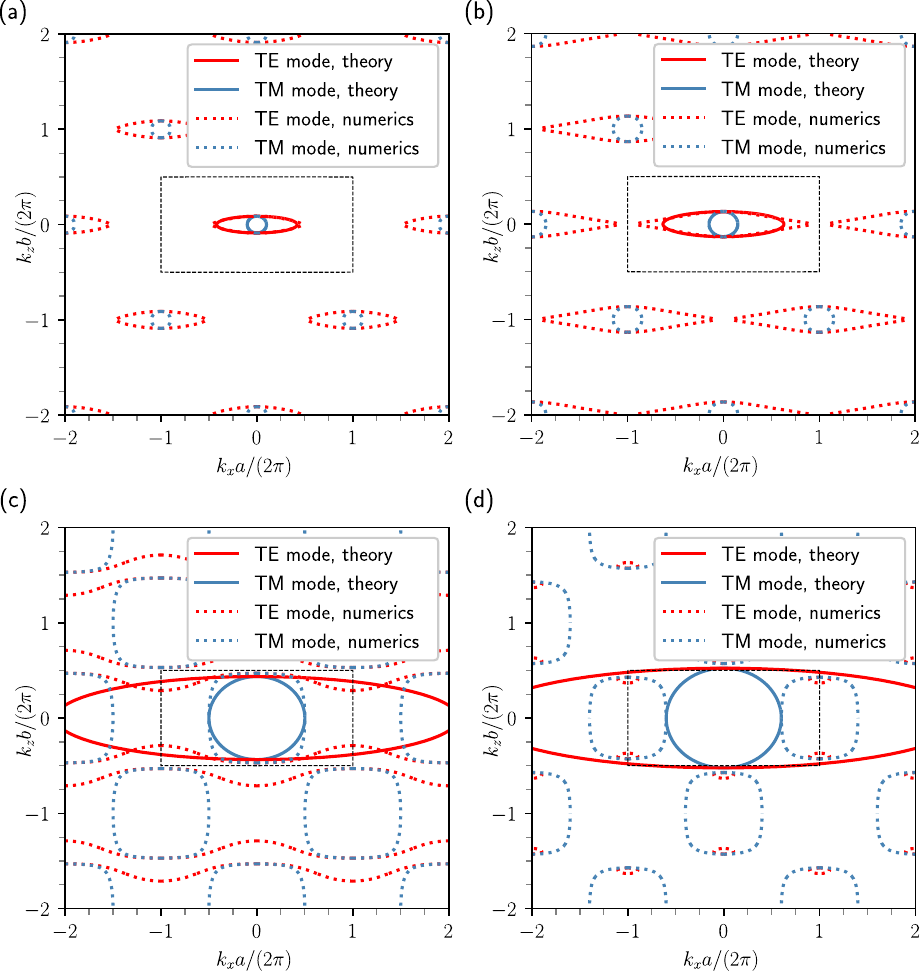}%
}
\caption{Theoretical and computational isofrequency contours for the square patch metamaterial in plane $k_y=0$ for normalized frequencies (a) $\omega a/(2\pi c)=0.10$, (b) $\omega a/(2\pi c)=0.15$, (c) $\omega a/(2\pi c)=0.50$, and (d) $\omega a/(2\pi c)=0.60$. The dotted black lines indicate the edges of the first Brillouin zone.}\label{fig_7}
\end{figure}
Following the dispersion analysis of the structure, isofrequency contours in plane $k_y=0$ are illustrated in Fig. \ref{fig_7}, in which the black dotted line indicates the edges of the first Brillouin zone. At low frequencies, i.e. $\omega a/(2\pi c)=0.10$, Fig. \hyperref[fig_7]{7a} shows a perfect agreement between theory and simulations for both TE and TM modes. However, this agreement weakens at higher frequencies. At $\omega a/(2\pi c)=0.15$, illustrated in Fig. \hyperref[fig_7]{7b}, it is noted that there is a perfect agreement for the TM mode, but the TE mode shows a discrepancy between theoretical and numerical results along directions close to $k_x$-direction. At $\omega a/(2\pi c)=0.50$, the agreement for the TE mode is limited to $k_z$-direction, and the agreement for TM mode is limited to $k_z$- and $k_x$- directions, see Fig. \hyperref[fig_7]{7c}. Finally, Fig. \hyperref[fig_7]{7d} shows a significant discrepancy at $\omega a/(2\pi c)=0.60$, indicating a breakdown of the model's validity due to a topological transition in the isofrequency contours \cite{krishnamoorthy2012topological}.

\subsection{Electromagnetic parameters dependence on geometry}
\begin{figure}[ht]
  {\includegraphics[width=1\columnwidth]{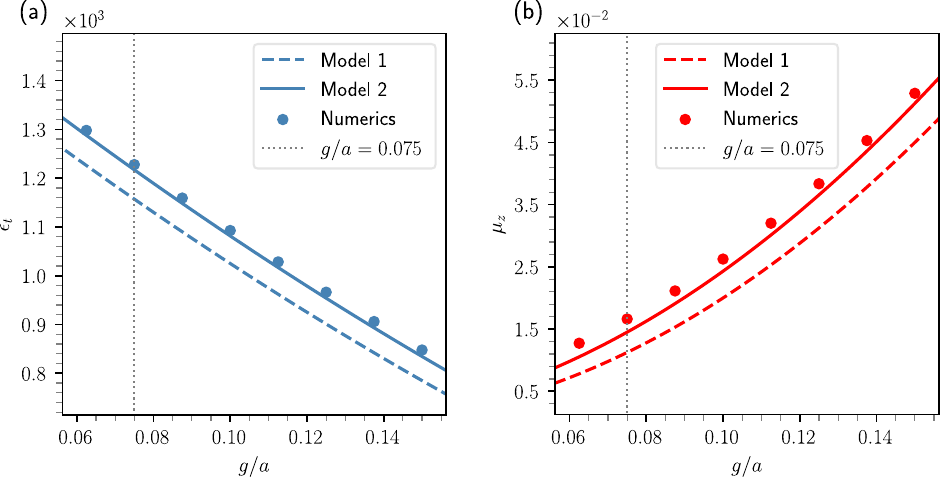}
  }
\caption{Effect of the gap size on (a) relative permittivity and (b) relative permeability of the  square patch metamaterial. Results are shown as a function of the gap size per transversal period ($g/a$), according to Model 1 (without edge effects), Model 2 (with edge effects), and numerical dispersion relations at low frequencies. The parameter $g/a=0.075$ was used to obtain the results shown in Sec. II-A.}\label{fig_8}
\end{figure}

In order to investigate the performance of the model under variations of the gap ($g$) between metallic patches, Fig. \ref{fig_8} shows theoretical and simulations results for the effective permittivity and permeability of the structure as a function of $g/a$ when $b/a=0.025$. The theoretical results are obtained directly from equations \eqref{eq_3ep1} and \eqref{eq_3mu1} for Model 1, and from equations \eqref{eq_3ep} and \eqref{eq_3mu} for Model 2, which include edge effects. On the other hand, numerical results are obtained from the dispersion along $k_z$-direction to retrieve first the transversal permittivity, considering that a linear dispersion is evidenced in Fig. \hyperref[fig_5]{5b} (path $\Gamma \mathrm{Z}$) at low frequencies,
\begin{equation}
    \epsilon_t=\left(k_z\frac{c}{\omega}\right)^2,
\end{equation}
where $k_z=\tfrac{\pi}{20b}$ to ensure we are sufficiently far from the edges of the Brillouin zone. Then, the axial permeability $(\mu_z)$ is retrieved from the dispersion along $k_x$-direction, considering a linear dispersion at low frequencies (see Fig. \hyperref[fig_5]{5a} , path $\Gamma \mathrm{X}$),

\begin{equation}
    \mu_z=\frac{1}{\epsilon_t}\left(k_x\frac{c}{\omega}\right)^2,
\end{equation}
where $k_x=\tfrac{\pi}{20a}$ to ensure we are sufficiently far from the edges of the Brillouin zone.

Fig. \hyperref[fig_8]{8a} illustrates the effective relative permittivity ($\epsilon_t$), showing an advantageous range of values from approximately $800$ to $1300$ for the interval $0.06\leq g/a\leq 0.15$. Similarly, Fig. \hyperref[fig_8]{8b} demonstrates a range of effective relative permeability values ($\mu_z$) from approximately $0.010$ to $0.050$ within the same interval. It is worth noting that Model 2 exhibits improved agreement with the numerical results compared to Model 1. A smaller $g/a$ ratio reduces the size of the vacuum canals and weakens the assumption of an exponential decay for the edge effect in the magnetic problem (see Supplemental Material \cite{SM}), leading to a larger relative error in the calculated magnetic permeability.

\begin{figure}[ht]
  {\includegraphics[width=1\columnwidth]{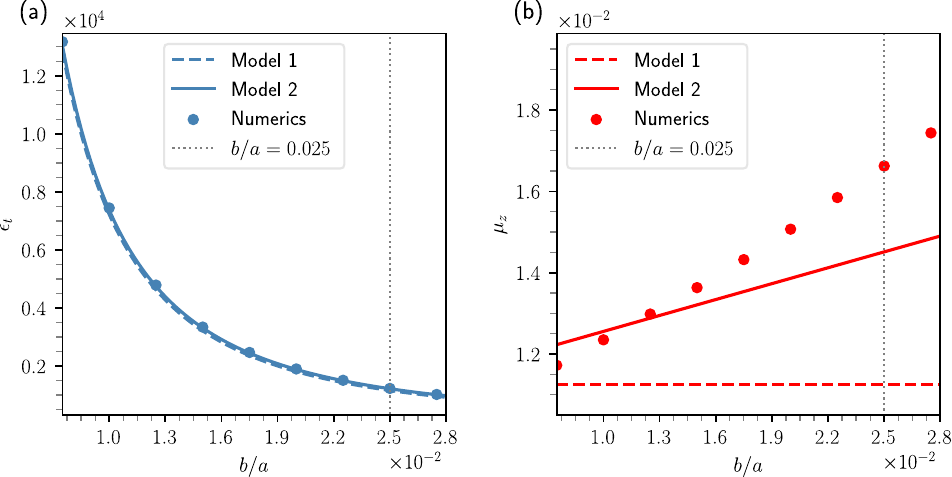}
  }
\caption{Effect of axial period on (a) relative permittivity and (b) relative permeability of the  square patch metamaterial. Results are shown as a function of the axial period per transversal period ($b/a$), according to Model 1 (without edge effects), Model 2 (with edge effects), and numerical dispersion relations at low frequencies. The parameter $b/a=0.025$ was used to obtain the results shown in Sec. II-A.}\label{fig_9}
\end{figure}

Figure \ref{fig_9} presents theoretical and simulation results, obtained using the aforementioned methodology, for the effective relative permittivity and effective relative permeability of the structure as a function of $b/a$ when $g/a=0.075$. Within the domain $0.0075\leq b/a\leq 0.028$ a wide range of transversal permittivity values are achieved in Fig. \hyperref[fig_9]{9a} (from approximately $1200$ to $12000$). In contrast, Fig. \hyperref[fig_9]{9b} shows minimal variation in permeability within the same domain, with values ranging from approximately $0.010$ to $0.017$. Notably, Model 2 exhibits significantly improved agreement with the numerical results compared to Model 1. When the $b/a$ ratio increases, the agreement of the theoretical models decreases with respect to numerical results since  the assumption of tightly packet patches weakens.

\section{Conclusions}

A theoretical and computational study was conducted to elucidate the dispersion characteristics of the square patch metamaterial. Our findings reveal that the structure emulates the behavior of a uniaxial dielectric magnetic material at low frequencies. Furthermore, we have established accurate and straightforward theoretical formulas for estimating the key electromagnetic parameters of these metamaterials. These formulas unveil the critical dependence of both the transversal permittivity and the axial permeability on the spatial periods and the gaps between the constituent patches. Moreover, the derived expressions predict that the square patch metamaterial not only exhibits an exceptionally large transverse permittivity but also possesses an extremely small axial permeability.

Comparing theoretical results with computational simulations based on a body-centered tetragonal representation of the square patch metamaterial, the theoretical model demonstrates its efficacy in predicting the dispersion characteristics up to approximately $\omega a/(2\pi c)=0.5$ coinciding with the onset of topological transitions. Notably, good agreement is observed between theory and simulations, with a relative error of $7.1\%$ at $\mathrm{X/3}$ point and $8.3\%$ at $\mathrm{Z}$ point. Furthermore, the theoretical formulas allow for the prediction of the relative transverse permittivity within a range of 800 to 12000, and the relative permeability within a range of approximately 0.010 to 0.050. These findings highlight the potential of the developed theoretical model for guiding the design of square patch metamaterials with tailored electromagnetic properties. Further research could explore the application of this model to similar structures with even more degrees of freedom, such as glide-symmetric structures with metallic objects with shapes different to the square shape, and investigate the influence of fabrication tolerances on the predicted properties. This would significantly advance the development of practical metamaterial devices, enabling functionalities such as angle-selective surfaces with even finer control over electromagnetic wave propagation or impedance matching layers with broader bandwidths.

For future research, we propose analyzing the losses of the structure to facilitate a experimental implementation. A practical finite sample of the square patch metamaterial could be realized using stacked printed circuit boards (PCBs). This change in host medium from air to the PCB dielectric would affect the effective relative permittivity of the metamaterial. As evident in Eqs. \eqref{eq:Ex} and \eqref{eq: et1}, the effective relative permittivity would be directly proportional to the relative permittivity of the host medium. However, this also implies that the host dielectric's loss tangent would be transferred to the structure. In addition, considering the finite conductivity of real metals and the typical thickness of metals in the microwave regime (often exceeding the skin depth), the skin effect approximation can be employed to estimate the microscopic electromagnetic fields within the metals and quantify the losses associated with the metal patches in the square patch metamaterial. An alternative approach is to consider a surface impedance to represent the metals with finite conductivity, as was used in \cite{8457250} to calculate the scattering parameters of align and non-align metal patches including losses. When the skin depth of the metal patches exceeds their thickness, the complex permittivity of the metal should be considered to study the microscopic fields within the metal, as demonstrated in \cite{chang_broadband_2016}.

\section{Acknowledgment}
The numerical simulations and the theoretical model that considers uniform surface charge density and uniform microscopic magnetic flux density were supported by the Ministry of Science and Higher Education of the Russian Federation (Project No. 075-15-2022-1120). The theoretical investigation of edge effects, including Supplemental Material \cite{SM} and the analysis on the dependence of electromagnetic parameters on geometry, was supported by State Assignment FSER-2024-0041 within the framework of the national project ``Science and Universities".
\providecommand{\noopsort}[1]{}\providecommand{\singleletter}[1]{#1}%

\begin{thebibliography}{32}%
\makeatletter
\providecommand \@ifxundefined [1]{%
 \@ifx{#1\undefined}
}%
\providecommand \@ifnum [1]{%
 \ifnum #1\expandafter \@firstoftwo
 \else \expandafter \@secondoftwo
 \fi
}%
\providecommand \@ifx [1]{%
 \ifx #1\expandafter \@firstoftwo
 \else \expandafter \@secondoftwo
 \fi
}%
\providecommand \natexlab [1]{#1}%
\providecommand \enquote  [1]{``#1''}%
\providecommand \bibnamefont  [1]{#1}%
\providecommand \bibfnamefont [1]{#1}%
\providecommand \citenamefont [1]{#1}%
\providecommand \href@noop [0]{\@secondoftwo}%
\providecommand \href [0]{\begingroup \@sanitize@url \@href}%
\providecommand \@href[1]{\@@startlink{#1}\@@href}%
\providecommand \@@href[1]{\endgroup#1\@@endlink}%
\providecommand \@sanitize@url [0]{\catcode `\\12\catcode `\$12\catcode `\&12\catcode `\#12\catcode `\^12\catcode `\_12\catcode `\%12\relax}%
\providecommand \@@startlink[1]{}%
\providecommand \@@endlink[0]{}%
\providecommand \url  [0]{\begingroup\@sanitize@url \@url }%
\providecommand \@url [1]{\endgroup\@href {#1}{\urlprefix }}%
\providecommand \urlprefix  [0]{URL }%
\providecommand \Eprint [0]{\href }%
\providecommand \doibase [0]{https://doi.org/}%
\providecommand \selectlanguage [0]{\@gobble}%
\providecommand \bibinfo  [0]{\@secondoftwo}%
\providecommand \bibfield  [0]{\@secondoftwo}%
\providecommand \translation [1]{[#1]}%
\providecommand \BibitemOpen [0]{}%
\providecommand \bibitemStop [0]{}%
\providecommand \bibitemNoStop [0]{.\EOS\space}%
\providecommand \EOS [0]{\spacefactor3000\relax}%
\providecommand \BibitemShut  [1]{\csname bibitem#1\endcsname}%
\let\auto@bib@innerbib\@empty
\bibitem [{\citenamefont {Poddubny}\ \emph {et~al.}(2013)\citenamefont {Poddubny}, \citenamefont {Iorsh}, \citenamefont {Belov},\ and\ \citenamefont {Kivshar}}]{poddubny_hyperbolic_2013}%
  \BibitemOpen
  \bibfield  {author} {\bibinfo {author} {\bibfnamefont {A.}~\bibnamefont {Poddubny}}, \bibinfo {author} {\bibfnamefont {I.}~\bibnamefont {Iorsh}}, \bibinfo {author} {\bibfnamefont {P.}~\bibnamefont {Belov}},\ and\ \bibinfo {author} {\bibfnamefont {Y.}~\bibnamefont {Kivshar}},\ }\bibfield  {title} {\bibinfo {title} {Hyperbolic metamaterials},\ }\href {https://doi.org/10.1038/nphoton.2013.243} {\bibfield  {journal} {\bibinfo  {journal} {Nature Photonics}\ }\textbf {\bibinfo {volume} {7}},\ \bibinfo {pages} {948} (\bibinfo {year} {2013})}\BibitemShut {NoStop}%
\bibitem [{\citenamefont {Zhang}\ and\ \citenamefont {Wu}(2015)}]{zhang_effective_2015}%
  \BibitemOpen
  \bibfield  {author} {\bibinfo {author} {\bibfnamefont {X.}~\bibnamefont {Zhang}}\ and\ \bibinfo {author} {\bibfnamefont {Y.}~\bibnamefont {Wu}},\ }\bibfield  {title} {\bibinfo {title} {Effective medium theory for anisotropic metamaterials},\ }\href {https://doi.org/10.1038/srep07892} {\bibfield  {journal} {\bibinfo  {journal} {Scientific Reports}\ }\textbf {\bibinfo {volume} {5}},\ \bibinfo {pages} {7892} (\bibinfo {year} {2015})}\BibitemShut {NoStop}%
\bibitem [{\citenamefont {Hao}\ \emph {et~al.}(2007)\citenamefont {Hao}, \citenamefont {Yuan}, \citenamefont {Ran}, \citenamefont {Jiang}, \citenamefont {Kong}, \citenamefont {Chan},\ and\ \citenamefont {Zhou}}]{hao_manipulating_2007}%
  \BibitemOpen
  \bibfield  {author} {\bibinfo {author} {\bibfnamefont {J.}~\bibnamefont {Hao}}, \bibinfo {author} {\bibfnamefont {Y.}~\bibnamefont {Yuan}}, \bibinfo {author} {\bibfnamefont {L.}~\bibnamefont {Ran}}, \bibinfo {author} {\bibfnamefont {T.}~\bibnamefont {Jiang}}, \bibinfo {author} {\bibfnamefont {J.~A.}\ \bibnamefont {Kong}}, \bibinfo {author} {\bibfnamefont {C.~T.}\ \bibnamefont {Chan}},\ and\ \bibinfo {author} {\bibfnamefont {L.}~\bibnamefont {Zhou}},\ }\bibfield  {title} {\bibinfo {title} {Manipulating {Electromagnetic} {Wave} {Polarizations} by {Anisotropic} {Metamaterials}},\ }\href {https://doi.org/10.1103/PhysRevLett.99.063908} {\bibfield  {journal} {\bibinfo  {journal} {Physical Review Letters}\ }\textbf {\bibinfo {volume} {99}},\ \bibinfo {pages} {063908} (\bibinfo {year} {2007})}\BibitemShut {NoStop}%
\bibitem [{\citenamefont {Hessel}\ \emph {et~al.}(1973)\citenamefont {Hessel}, \citenamefont {Chen}, \citenamefont {Li},\ and\ \citenamefont {Oliner}}]{1450933}%
  \BibitemOpen
  \bibfield  {author} {\bibinfo {author} {\bibfnamefont {A.}~\bibnamefont {Hessel}}, \bibinfo {author} {\bibfnamefont {M.~H.}\ \bibnamefont {Chen}}, \bibinfo {author} {\bibfnamefont {R.}~\bibnamefont {Li}},\ and\ \bibinfo {author} {\bibfnamefont {A.}~\bibnamefont {Oliner}},\ }\bibfield  {title} {\bibinfo {title} {Propagation in periodically loaded waveguides with higher symmetries},\ }\href {https://doi.org/10.1109/PROC.1973.9003} {\bibfield  {journal} {\bibinfo  {journal} {Proceedings of the IEEE}\ }\textbf {\bibinfo {volume} {61}},\ \bibinfo {pages} {183} (\bibinfo {year} {1973})}\BibitemShut {NoStop}%
\bibitem [{\citenamefont {Quevedo-Teruel}\ \emph {et~al.}(2020)\citenamefont {Quevedo-Teruel}, \citenamefont {Valerio}, \citenamefont {Sipus},\ and\ \citenamefont {Rajo-Iglesias}}]{quevedo-teruel_periodic_2020}%
  \BibitemOpen
  \bibfield  {author} {\bibinfo {author} {\bibfnamefont {O.}~\bibnamefont {Quevedo-Teruel}}, \bibinfo {author} {\bibfnamefont {G.}~\bibnamefont {Valerio}}, \bibinfo {author} {\bibfnamefont {Z.}~\bibnamefont {Sipus}},\ and\ \bibinfo {author} {\bibfnamefont {E.}~\bibnamefont {Rajo-Iglesias}},\ }\bibfield  {title} {\bibinfo {title} {Periodic {Structures} {With} {Higher} {Symmetries}: {Their} {Applications} in {Electromagnetic} {Devices}},\ }\href {https://doi.org/10.1109/MMM.2020.3014987} {\bibfield  {journal} {\bibinfo  {journal} {IEEE Microwave Magazine}\ }\textbf {\bibinfo {volume} {21}},\ \bibinfo {pages} {36} (\bibinfo {year} {2020})}\BibitemShut {NoStop}%
\bibitem [{\citenamefont {Quevedo-Teruel}\ \emph {et~al.}(2021)\citenamefont {Quevedo-Teruel}, \citenamefont {Chen}, \citenamefont {Mesa}, \citenamefont {Fonseca},\ and\ \citenamefont {Valerio}}]{quevedo-teruel_benefits_2021}%
  \BibitemOpen
  \bibfield  {author} {\bibinfo {author} {\bibfnamefont {O.}~\bibnamefont {Quevedo-Teruel}}, \bibinfo {author} {\bibfnamefont {Q.}~\bibnamefont {Chen}}, \bibinfo {author} {\bibfnamefont {F.}~\bibnamefont {Mesa}}, \bibinfo {author} {\bibfnamefont {N.~J.}\ \bibnamefont {Fonseca}},\ and\ \bibinfo {author} {\bibfnamefont {G.}~\bibnamefont {Valerio}},\ }\bibfield  {title} {\bibinfo {title} {On the {Benefits} of {Glide} {Symmetries} for {Microwave} {Devices}},\ }\href {https://doi.org/10.1109/JMW.2020.3033847} {\bibfield  {journal} {\bibinfo  {journal} {IEEE Journal of Microwaves}\ }\textbf {\bibinfo {volume} {1}},\ \bibinfo {pages} {457} (\bibinfo {year} {2021})}\BibitemShut {NoStop}%
\bibitem [{\citenamefont {Ebrahimpouri}\ and\ \citenamefont {Quevedo-Teruel}(2019)}]{ebrahimpouri_ultrawideband_2019}%
  \BibitemOpen
  \bibfield  {author} {\bibinfo {author} {\bibfnamefont {M.}~\bibnamefont {Ebrahimpouri}}\ and\ \bibinfo {author} {\bibfnamefont {O.}~\bibnamefont {Quevedo-Teruel}},\ }\bibfield  {title} {\bibinfo {title} {Ultrawideband {Anisotropic} {Glide}-{Symmetric} {Metasurfaces}},\ }\href {https://doi.org/10.1109/LAWP.2019.2922238} {\bibfield  {journal} {\bibinfo  {journal} {IEEE Antennas and Wireless Propagation Letters}\ }\textbf {\bibinfo {volume} {18}},\ \bibinfo {pages} {1547} (\bibinfo {year} {2019})}\BibitemShut {NoStop}%
\bibitem [{\citenamefont {Arnberg}\ \emph {et~al.}(2020)\citenamefont {Arnberg}, \citenamefont {Barreira~Petersson}, \citenamefont {Zetterstrom}, \citenamefont {Ghasemifard},\ and\ \citenamefont {Quevedo-Teruel}}]{arnberg_high_2020}%
  \BibitemOpen
  \bibfield  {author} {\bibinfo {author} {\bibfnamefont {P.}~\bibnamefont {Arnberg}}, \bibinfo {author} {\bibfnamefont {O.}~\bibnamefont {Barreira~Petersson}}, \bibinfo {author} {\bibfnamefont {O.}~\bibnamefont {Zetterstrom}}, \bibinfo {author} {\bibfnamefont {F.}~\bibnamefont {Ghasemifard}},\ and\ \bibinfo {author} {\bibfnamefont {O.}~\bibnamefont {Quevedo-Teruel}},\ }\bibfield  {title} {\bibinfo {title} {High {Refractive} {Index} {Electromagnetic} {Devices} in {Printed} {Technology} {Based} on {Glide}-{Symmetric} {Periodic} {Structures}},\ }\href {https://doi.org/10.3390/app10093216} {\bibfield  {journal} {\bibinfo  {journal} {Applied Sciences}\ }\textbf {\bibinfo {volume} {10}},\ \bibinfo {pages} {3216} (\bibinfo {year} {2020})}\BibitemShut {NoStop}%
\bibitem [{\citenamefont {Chang}\ \emph {et~al.}(2016)\citenamefont {Chang}, \citenamefont {Kim}, \citenamefont {Kang}, \citenamefont {Kim}, \citenamefont {Kim}, \citenamefont {Lee},\ and\ \citenamefont {Shin}}]{chang_broadband_2016}%
  \BibitemOpen
  \bibfield  {author} {\bibinfo {author} {\bibfnamefont {T.}~\bibnamefont {Chang}}, \bibinfo {author} {\bibfnamefont {J.~U.}\ \bibnamefont {Kim}}, \bibinfo {author} {\bibfnamefont {S.~K.}\ \bibnamefont {Kang}}, \bibinfo {author} {\bibfnamefont {H.}~\bibnamefont {Kim}}, \bibinfo {author} {\bibfnamefont {D.~K.}\ \bibnamefont {Kim}}, \bibinfo {author} {\bibfnamefont {Y.-H.}\ \bibnamefont {Lee}},\ and\ \bibinfo {author} {\bibfnamefont {J.}~\bibnamefont {Shin}},\ }\bibfield  {title} {\bibinfo {title} {Broadband giant-refractive-index material based on mesoscopic space-filling curves},\ }\href {https://doi.org/10.1038/ncomms12661} {\bibfield  {journal} {\bibinfo  {journal} {Nature Communications}\ }\textbf {\bibinfo {volume} {7}},\ \bibinfo {pages} {12661} (\bibinfo {year} {2016})}\BibitemShut {NoStop}%
\bibitem [{\citenamefont {Barzegar-Parizi}\ and\ \citenamefont {Rejaei}(2015)}]{barzegar-parizi_calculation_2015}%
  \BibitemOpen
  \bibfield  {author} {\bibinfo {author} {\bibfnamefont {S.}~\bibnamefont {Barzegar-Parizi}}\ and\ \bibinfo {author} {\bibfnamefont {B.}~\bibnamefont {Rejaei}},\ }\bibfield  {title} {\bibinfo {title} {Calculation of effective parameters of high permittivity integrated artificial dielectrics},\ }\href {https://doi.org/10.1049/iet-map.2014.0377} {\bibfield  {journal} {\bibinfo  {journal} {IET Microwaves, Antennas \& Propagation}\ }\textbf {\bibinfo {volume} {9}},\ \bibinfo {pages} {1287} (\bibinfo {year} {2015})}\BibitemShut {NoStop}%
\bibitem [{\citenamefont {Ma}\ \emph {et~al.}(2008)\citenamefont {Ma}, \citenamefont {Rejaei},\ and\ \citenamefont {Zhuang}}]{ma_artificial_2008}%
  \BibitemOpen
  \bibfield  {author} {\bibinfo {author} {\bibfnamefont {Y.}~\bibnamefont {Ma}}, \bibinfo {author} {\bibfnamefont {B.}~\bibnamefont {Rejaei}},\ and\ \bibinfo {author} {\bibfnamefont {Y.}~\bibnamefont {Zhuang}},\ }\bibfield  {title} {\bibinfo {title} {Artificial {Dielectric} {Shields} {forIntegrated} {Transmission} {Lines}},\ }\href {https://doi.org/10.1109/LMWC.2008.924907} {\bibfield  {journal} {\bibinfo  {journal} {IEEE Microwave and Wireless Components Letters}\ }\textbf {\bibinfo {volume} {18}},\ \bibinfo {pages} {431} (\bibinfo {year} {2008})}\BibitemShut {NoStop}%
\bibitem [{\citenamefont {Takahagi}\ and\ \citenamefont {Sano}(2011)}]{takahagi_high-gain_2011}%
  \BibitemOpen
  \bibfield  {author} {\bibinfo {author} {\bibfnamefont {K.}~\bibnamefont {Takahagi}}\ and\ \bibinfo {author} {\bibfnamefont {E.}~\bibnamefont {Sano}},\ }\bibfield  {title} {\bibinfo {title} {High-{Gain} {Silicon} {On}-{Chip} {Antenna} {With} {Artificial} {Dielectric} {Layer}},\ }\href {https://doi.org/10.1109/TAP.2011.2163758} {\bibfield  {journal} {\bibinfo  {journal} {IEEE Transactions on Antennas and Propagation}\ }\textbf {\bibinfo {volume} {59}},\ \bibinfo {pages} {3624} (\bibinfo {year} {2011})}\BibitemShut {NoStop}%
\bibitem [{\citenamefont {Vorobyev}\ \emph {et~al.}(2020)\citenamefont {Vorobyev}, \citenamefont {Shchelokova}, \citenamefont {Zivkovic}, \citenamefont {Slobozhanyuk}, \citenamefont {Baena}, \citenamefont {del Risco}, \citenamefont {Abdeddaim}, \citenamefont {Webb},\ and\ \citenamefont {Glybovski}}]{vorobyev_artificial_2020}%
  \BibitemOpen
  \bibfield  {author} {\bibinfo {author} {\bibfnamefont {V.}~\bibnamefont {Vorobyev}}, \bibinfo {author} {\bibfnamefont {A.}~\bibnamefont {Shchelokova}}, \bibinfo {author} {\bibfnamefont {I.}~\bibnamefont {Zivkovic}}, \bibinfo {author} {\bibfnamefont {A.}~\bibnamefont {Slobozhanyuk}}, \bibinfo {author} {\bibfnamefont {J.~D.}\ \bibnamefont {Baena}}, \bibinfo {author} {\bibfnamefont {J.~P.}\ \bibnamefont {del Risco}}, \bibinfo {author} {\bibfnamefont {R.}~\bibnamefont {Abdeddaim}}, \bibinfo {author} {\bibfnamefont {A.}~\bibnamefont {Webb}},\ and\ \bibinfo {author} {\bibfnamefont {S.}~\bibnamefont {Glybovski}},\ }\bibfield  {title} {\bibinfo {title} {An artificial dielectric slab for ultra high-field {MRI}: {Proof} of concept},\ }\href {https://doi.org/10.1016/j.jmr.2020.106835} {\bibfield  {journal} {\bibinfo  {journal} {Journal of Magnetic Resonance}\ }\textbf {\bibinfo {volume} {320}},\ \bibinfo {pages} {106835} (\bibinfo {year} {2020})}\BibitemShut {NoStop}%
\bibitem [{\citenamefont {Cavallo}\ and\ \citenamefont {Felita}(2017)}]{cavallo_analytical_2017}%
  \BibitemOpen
  \bibfield  {author} {\bibinfo {author} {\bibfnamefont {D.}~\bibnamefont {Cavallo}}\ and\ \bibinfo {author} {\bibfnamefont {C.}~\bibnamefont {Felita}},\ }\bibfield  {title} {\bibinfo {title} {Analytical {Formulas} for {Artificial} {Dielectrics} {With} {Nonaligned} {Layers}},\ }\href {https://doi.org/10.1109/TAP.2017.2738064} {\bibfield  {journal} {\bibinfo  {journal} {IEEE Transactions on Antennas and Propagation}\ }\textbf {\bibinfo {volume} {65}},\ \bibinfo {pages} {5303} (\bibinfo {year} {2017})}\BibitemShut {NoStop}%
\bibitem [{\citenamefont {Cavallo}(2018)}]{8457250}%
  \BibitemOpen
  \bibfield  {author} {\bibinfo {author} {\bibfnamefont {D.}~\bibnamefont {Cavallo}},\ }\bibfield  {title} {\bibinfo {title} {Dissipation losses in artificial dielectric layers},\ }\href {https://doi.org/10.1109/TAP.2018.2869241} {\bibfield  {journal} {\bibinfo  {journal} {IEEE Transactions on Antennas and Propagation}\ }\textbf {\bibinfo {volume} {66}},\ \bibinfo {pages} {7460} (\bibinfo {year} {2018})}\BibitemShut {NoStop}%
\bibitem [{\citenamefont {Baena}\ \emph {et~al.}(2020)\citenamefont {Baena}, \citenamefont {del Risco},\ and\ \citenamefont {Escobar}}]{Baena_boradband_uniaxial2020}%
  \BibitemOpen
  \bibfield  {author} {\bibinfo {author} {\bibfnamefont {J.~D.}\ \bibnamefont {Baena}}, \bibinfo {author} {\bibfnamefont {J.~P.}\ \bibnamefont {del Risco}},\ and\ \bibinfo {author} {\bibfnamefont {A.~C.}\ \bibnamefont {Escobar}},\ }\bibfield  {title} {\bibinfo {title} {Broadband uniaxial dielectric-magnetic metamaterial with giant anisotropy factor},\ }in\ \href {https://doi.org/10.1109/Metamaterials49557.2020.9284987} {\emph {\bibinfo {booktitle} {2020 Fourteenth International Congress on Artificial Materials for Novel Wave Phenomena (Metamaterials)}}}\ (\bibinfo {year} {2020})\ pp.\ \bibinfo {pages} {367--369}\BibitemShut {NoStop}%
\bibitem [{\citenamefont {Tretyakov}(2003)}]{tretyakov2003analytical}%
  \BibitemOpen
  \bibfield  {author} {\bibinfo {author} {\bibfnamefont {S.}~\bibnamefont {Tretyakov}},\ }\href@noop {} {\emph {\bibinfo {title} {Analytical modeling in applied electromagnetics}}}\ (\bibinfo  {publisher} {Artech House},\ \bibinfo {year} {2003})\BibitemShut {NoStop}%
\bibitem [{SM()}]{SM}%
  \BibitemOpen
  \href@noop {} {\bibinfo {title} {See supplemental material at [url will be inserted by publisher] for the derivation of edge effect correction for permittivity and permeability formulas, which includes ref. [19]}}\BibitemShut {NoStop}%
\bibitem [{\citenamefont {Pozar}(2012)}]{Pozar:882338}%
  \BibitemOpen
  \bibfield  {author} {\bibinfo {author} {\bibfnamefont {D.~M.}\ \bibnamefont {Pozar}},\ }\href@noop {} {\emph {\bibinfo {title} {Microwave engineering}}},\ \bibinfo {edition} {4th}\ ed.,\ \bibinfo {series} {Wiley Series in Microwave and Optical Engineering}, Vol.~\bibinfo {volume} {2}\ (\bibinfo  {publisher} {Wiley},\ \bibinfo {year} {2012})\BibitemShut {NoStop}%
\bibitem [{\citenamefont {del Risco}\ and\ \citenamefont {Baena}(2016)}]{7746420}%
  \BibitemOpen
  \bibfield  {author} {\bibinfo {author} {\bibfnamefont {J.~P.}\ \bibnamefont {del Risco}}\ and\ \bibinfo {author} {\bibfnamefont {J.~D.}\ \bibnamefont {Baena}},\ }\bibfield  {title} {\bibinfo {title} {Extremely thin fabry-perot resonators based on high permitivity artificial dielectric},\ }in\ \href {https://doi.org/10.1109/MetaMaterials.2016.7746420} {\emph {\bibinfo {booktitle} {2016 10th International Congress on Advanced Electromagnetic Materials in Microwaves and Optics (METAMATERIALS)}}}\ (\bibinfo {year} {2016})\ pp.\ \bibinfo {pages} {40--42}\BibitemShut {NoStop}%
\bibitem [{\citenamefont {Depine}\ \emph {et~al.}(2006)\citenamefont {Depine}, \citenamefont {Inchaussandague},\ and\ \citenamefont {Lakhtakia}}]{depine_2006}%
  \BibitemOpen
  \bibfield  {author} {\bibinfo {author} {\bibfnamefont {R.~A.}\ \bibnamefont {Depine}}, \bibinfo {author} {\bibfnamefont {M.~E.}\ \bibnamefont {Inchaussandague}},\ and\ \bibinfo {author} {\bibfnamefont {A.}~\bibnamefont {Lakhtakia}},\ }\bibfield  {title} {\bibinfo {title} {Classification of dispersion equations for homogeneous, dielectric-magnetic, uniaxial materials},\ }\href {https://doi.org/10.1364/JOSAA.23.000949} {\bibfield  {journal} {\bibinfo  {journal} {JOSA A}\ }\textbf {\bibinfo {volume} {23}},\ \bibinfo {pages} {949} (\bibinfo {year} {2006})}\BibitemShut {NoStop}%
\bibitem [{\citenamefont {Lakhtakia}\ and\ \citenamefont {Weiglhofer}(1997)}]{lakhtakia1997time}%
  \BibitemOpen
  \bibfield  {author} {\bibinfo {author} {\bibfnamefont {A.}~\bibnamefont {Lakhtakia}}\ and\ \bibinfo {author} {\bibfnamefont {W.~S.}\ \bibnamefont {Weiglhofer}},\ }\bibfield  {title} {\bibinfo {title} {Time-harmonic electromagnetic field in a source region in a uniaxial dielectric-magnetic medium},\ }\href@noop {} {\bibfield  {journal} {\bibinfo  {journal} {International Journal of Applied Electromagnetics and Mechanics}\ }\textbf {\bibinfo {volume} {8}},\ \bibinfo {pages} {167} (\bibinfo {year} {1997})}\BibitemShut {NoStop}%
\bibitem [{\citenamefont {Weiglhofer}\ and\ \citenamefont {Lakhtakia}(1996)}]{weiglhofer1996new}%
  \BibitemOpen
  \bibfield  {author} {\bibinfo {author} {\bibfnamefont {W.}~\bibnamefont {Weiglhofer}}\ and\ \bibinfo {author} {\bibfnamefont {A.}~\bibnamefont {Lakhtakia}},\ }\bibfield  {title} {\bibinfo {title} {New expressions for depolarization dyadics in uniaxial dielectric-magnetic media},\ }\href {https://doi.org/https://doi.org/10.1007/BF02068788} {\bibfield  {journal} {\bibinfo  {journal} {International journal of infrared and millimeter waves}\ }\textbf {\bibinfo {volume} {17}},\ \bibinfo {pages} {1365} (\bibinfo {year} {1996})}\BibitemShut {NoStop}%
\bibitem [{\citenamefont {Huang}\ and\ \citenamefont {Shen}(2020)}]{huang_angle-selective_2020}%
  \BibitemOpen
  \bibfield  {author} {\bibinfo {author} {\bibfnamefont {H.}~\bibnamefont {Huang}}\ and\ \bibinfo {author} {\bibfnamefont {Z.}~\bibnamefont {Shen}},\ }\bibfield  {title} {\bibinfo {title} {Angle-{Selective} {Surface} {Based} on {Uniaxial} {Dielectric}-{Magnetic} {Slab}},\ }\href {https://doi.org/10.1109/LAWP.2020.3035439} {\bibfield  {journal} {\bibinfo  {journal} {IEEE Antennas and Wireless Propagation Letters}\ }\textbf {\bibinfo {volume} {19}},\ \bibinfo {pages} {2457} (\bibinfo {year} {2020})}\BibitemShut {NoStop}%
\bibitem [{\citenamefont {He}\ and\ \citenamefont {Eleftheriades}(2018)}]{he_magnetoelectric_2018}%
  \BibitemOpen
  \bibfield  {author} {\bibinfo {author} {\bibfnamefont {Y.}~\bibnamefont {He}}\ and\ \bibinfo {author} {\bibfnamefont {G.~V.}\ \bibnamefont {Eleftheriades}},\ }\bibfield  {title} {\bibinfo {title} {Magnetoelectric uniaxial metamaterials as wide-angle polarization-insensitive matching layers},\ }\href {https://doi.org/10.1103/PhysRevB.98.205404} {\bibfield  {journal} {\bibinfo  {journal} {Phys. Rev. B}\ }\textbf {\bibinfo {volume} {98}},\ \bibinfo {pages} {205404} (\bibinfo {year} {2018})}\BibitemShut {NoStop}%
\bibitem [{\citenamefont {Sievenpiper}\ \emph {et~al.}(1999)\citenamefont {Sievenpiper}, \citenamefont {Zhang}, \citenamefont {Broas}, \citenamefont {Alexopolous},\ and\ \citenamefont {Yablonovitch}}]{798001}%
  \BibitemOpen
  \bibfield  {author} {\bibinfo {author} {\bibfnamefont {D.}~\bibnamefont {Sievenpiper}}, \bibinfo {author} {\bibfnamefont {L.}~\bibnamefont {Zhang}}, \bibinfo {author} {\bibfnamefont {R.}~\bibnamefont {Broas}}, \bibinfo {author} {\bibfnamefont {N.}~\bibnamefont {Alexopolous}},\ and\ \bibinfo {author} {\bibfnamefont {E.}~\bibnamefont {Yablonovitch}},\ }\bibfield  {title} {\bibinfo {title} {High-impedance electromagnetic surfaces with a forbidden frequency band},\ }\href {https://doi.org/10.1109/22.798001} {\bibfield  {journal} {\bibinfo  {journal} {IEEE Transactions on Microwave Theory and Techniques}\ }\textbf {\bibinfo {volume} {47}},\ \bibinfo {pages} {2059} (\bibinfo {year} {1999})}\BibitemShut {NoStop}%
\bibitem [{\citenamefont {Mouris}\ \emph {et~al.}(2020)\citenamefont {Mouris}, \citenamefont {Fernández-Prieto}, \citenamefont {Thobaben}, \citenamefont {Martel}, \citenamefont {Mesa},\ and\ \citenamefont {Quevedo-Teruel}}]{8972928}%
  \BibitemOpen
  \bibfield  {author} {\bibinfo {author} {\bibfnamefont {B.~A.}\ \bibnamefont {Mouris}}, \bibinfo {author} {\bibfnamefont {A.}~\bibnamefont {Fernández-Prieto}}, \bibinfo {author} {\bibfnamefont {R.}~\bibnamefont {Thobaben}}, \bibinfo {author} {\bibfnamefont {J.}~\bibnamefont {Martel}}, \bibinfo {author} {\bibfnamefont {F.}~\bibnamefont {Mesa}},\ and\ \bibinfo {author} {\bibfnamefont {O.}~\bibnamefont {Quevedo-Teruel}},\ }\bibfield  {title} {\bibinfo {title} {On the increment of the bandwidth of mushroom-type ebg structures with glide symmetry},\ }\href {https://doi.org/10.1109/TMTT.2020.2966700} {\bibfield  {journal} {\bibinfo  {journal} {IEEE Transactions on Microwave Theory and Techniques}\ }\textbf {\bibinfo {volume} {68}},\ \bibinfo {pages} {1365} (\bibinfo {year} {2020})}\BibitemShut {NoStop}%
\bibitem [{\citenamefont {Mehl}\ \emph {et~al.}(2017)\citenamefont {Mehl}, \citenamefont {Hicks}, \citenamefont {Toher}, \citenamefont {Levy}, \citenamefont {Hanson}, \citenamefont {Hart},\ and\ \citenamefont {Curtarolo}}]{MEHL2017S1}%
  \BibitemOpen
  \bibfield  {author} {\bibinfo {author} {\bibfnamefont {M.~J.}\ \bibnamefont {Mehl}}, \bibinfo {author} {\bibfnamefont {D.}~\bibnamefont {Hicks}}, \bibinfo {author} {\bibfnamefont {C.}~\bibnamefont {Toher}}, \bibinfo {author} {\bibfnamefont {O.}~\bibnamefont {Levy}}, \bibinfo {author} {\bibfnamefont {R.~M.}\ \bibnamefont {Hanson}}, \bibinfo {author} {\bibfnamefont {G.}~\bibnamefont {Hart}},\ and\ \bibinfo {author} {\bibfnamefont {S.}~\bibnamefont {Curtarolo}},\ }\bibfield  {title} {\bibinfo {title} {The aflow library of crystallographic prototypes: Part 1},\ }\href {https://doi.org/https://doi.org/10.1016/j.commatsci.2017.01.017} {\bibfield  {journal} {\bibinfo  {journal} {Computational Materials Science}\ }\textbf {\bibinfo {volume} {136}},\ \bibinfo {pages} {S1} (\bibinfo {year} {2017})}\BibitemShut {NoStop}%
\bibitem [{\citenamefont {Sakhno}\ \emph {et~al.}(2021)\citenamefont {Sakhno}, \citenamefont {Koreshin},\ and\ \citenamefont {Belov}}]{sakhno_longitudinal_2021}%
  \BibitemOpen
  \bibfield  {author} {\bibinfo {author} {\bibfnamefont {D.}~\bibnamefont {Sakhno}}, \bibinfo {author} {\bibfnamefont {E.}~\bibnamefont {Koreshin}},\ and\ \bibinfo {author} {\bibfnamefont {P.~A.}\ \bibnamefont {Belov}},\ }\bibfield  {title} {\bibinfo {title} {Longitudinal electromagnetic waves with extremely short wavelength},\ }\href {https://doi.org/10.1103/PhysRevB.104.L100304} {\bibfield  {journal} {\bibinfo  {journal} {Phys. Rev. B}\ }\textbf {\bibinfo {volume} {104}},\ \bibinfo {pages} {L100304} (\bibinfo {year} {2021})}\BibitemShut {NoStop}%
\bibitem [{\citenamefont {Yang}\ \emph {et~al.}(2023)\citenamefont {Yang}, \citenamefont {Zetterstrom}, \citenamefont {Mesa},\ and\ \citenamefont {Quevedo-Teruel}}]{yang2023dispersion}%
  \BibitemOpen
  \bibfield  {author} {\bibinfo {author} {\bibfnamefont {S.}~\bibnamefont {Yang}}, \bibinfo {author} {\bibfnamefont {O.}~\bibnamefont {Zetterstrom}}, \bibinfo {author} {\bibfnamefont {F.}~\bibnamefont {Mesa}},\ and\ \bibinfo {author} {\bibfnamefont {O.}~\bibnamefont {Quevedo-Teruel}},\ }\bibfield  {title} {\bibinfo {title} {Dispersion analysis of metasurfaces with hexagonal lattices with higher symmetries},\ }\href {https://ieeexplore.ieee.org/document/10256254} {\bibfield  {journal} {\bibinfo  {journal} {IEEE Journal of Microwaves}\ } (\bibinfo {year} {2023})}\BibitemShut {NoStop}%
\bibitem [{\citenamefont {Belov}\ and\ \citenamefont {Simovski}(2005)}]{Belov_homogenization_of_electromagnetic_crystals2005}%
  \BibitemOpen
  \bibfield  {author} {\bibinfo {author} {\bibfnamefont {P.~A.}\ \bibnamefont {Belov}}\ and\ \bibinfo {author} {\bibfnamefont {C.~R.}\ \bibnamefont {Simovski}},\ }\bibfield  {title} {\bibinfo {title} {Homogenization of electromagnetic crystals formed by uniaxial resonant scatterers},\ }\href {https://doi.org/10.1103/PhysRevE.72.026615} {\bibfield  {journal} {\bibinfo  {journal} {Phys. Rev. E}\ }\textbf {\bibinfo {volume} {72}},\ \bibinfo {pages} {026615} (\bibinfo {year} {2005})}\BibitemShut {NoStop}%
\bibitem [{\citenamefont {Krishnamoorthy}\ \emph {et~al.}(2012)\citenamefont {Krishnamoorthy}, \citenamefont {Jacob}, \citenamefont {Narimanov}, \citenamefont {Kretzschmar},\ and\ \citenamefont {Menon}}]{krishnamoorthy2012topological}%
  \BibitemOpen
  \bibfield  {author} {\bibinfo {author} {\bibfnamefont {H.~N.}\ \bibnamefont {Krishnamoorthy}}, \bibinfo {author} {\bibfnamefont {Z.}~\bibnamefont {Jacob}}, \bibinfo {author} {\bibfnamefont {E.}~\bibnamefont {Narimanov}}, \bibinfo {author} {\bibfnamefont {I.}~\bibnamefont {Kretzschmar}},\ and\ \bibinfo {author} {\bibfnamefont {V.~M.}\ \bibnamefont {Menon}},\ }\bibfield  {title} {\bibinfo {title} {Topological transitions in metamaterials},\ }\href {https://doi.org/10.1126/science.1219171} {\bibfield  {journal} {\bibinfo  {journal} {Science}\ }\textbf {\bibinfo {volume} {336}},\ \bibinfo {pages} {205} (\bibinfo {year} {2012})}\BibitemShut {NoStop}%
\end{thebibliography}

\providecommand{\noopsort}[1]{}\providecommand{\singleletter}[1]{#1}%

\end{document}